\documentclass[prb,twocolumn,showpacs,amsmath,amssymb]{revtex4}
\usepackage{graphicx,bm}

\newcommand{\E}{\mbox{e}}
\newcommand{\I}{\mbox{i}}
\newcommand{\SpringConstant}{\ensuremath{K}}
\newcommand{\EquilMass}{\ensuremath{m_{\circ}}}
\newcommand{\ImpurityMass}{\ensuremath{m_+}}
\newcommand{\UnitLength}{\ensuremath{a}}
\newcommand{\MassDensity}{\ensuremath{\mu}}
\newcommand{\YoungsModulus}{\ensuremath{E}}
\newcommand{\CrossSection}{\ensuremath{\sigma}}
\newcommand{\WaveNumber}{\ensuremath{k}}
\newcommand{\PulseSpeed}{\ensuremath{c_s}}
\newcommand{\Wavelength}{\lambda}
\newcommand{\Resistivity}{\rho}
\newcommand{\LocLength}{\xi}
\newcommand{\LocLengthDL}{\LocLength_\circ}	
\newcommand{\LExp}{\gamma}	
\newcommand{\MFP}{\Lambda}		

\newcommand{\SD}[1]{{s_{#1}}}    	
\newcommand{\VAR}[1]{{s^2_{#1}}}	
\newcommand{\Moment}{\Delta}		

\newcommand{\RNone}{\uppercase\expandafter{\romannumeral1}}
\newcommand{\RNtwo}{\uppercase\expandafter{\romannumeral2}}
\newcommand{\RNthree}{\uppercase\expandafter{\romannumeral3}}

\begin{document}
  \title{%
	Wave localization in binary isotopically disordered one-dimensional harmonic chains
		with impurities having arbitrary cross section and concentration
	}

  \author{K.A.~Snyder}
	\affiliation{Materials and Construction Division, 
		National Institute of Standards and Technology}

  \author{T.R.~Kirkpatrick}
	\affiliation{Institute for Physical Science and Technology and 
		Department of Physics, University of Maryland}

  \date{\today}


\begin{abstract}
The localization length for isotopically disordered harmonic one-dimensional
chains is calculated for arbitrary impurity 
concentration and scattering cross section.
The localization length depends on the
scattering cross section of a single scatterer, which is calculated for a 
discrete chain having a wavelength dependent pulse propagation 
speed.
%
%
%
For binary isotopically disordered systems composed of many scatterers, 
the localization length decreases with increasing impurity concentration, 
reaching a mimimum before diverging toward infinity as the impurity 
concentration approaches a value of one.
The concentration dependence of the localization length over the entire 
impurity concentration range is 
approximated accurately by the sum of the behavior at each limiting 
concentration.
Simultaneous 
measurements of Lyapunov exponent statistics indicate practical 
limits for the 
minimum system length and the number of scatterers to achieve representative 
ensemble averages.
Results are discussed in the context of 
future investigations of the time-dependent behavior of disordered 
anharmonic chains.
%
%
\end{abstract}

\maketitle

%
%

\section{Introduction}

The length scale over which Anderson localization \cite{Anderson58} occurs 
in harmonic disordered chains can be expressed as a function of the 
ensemble-averaged system resistivity 
\cite{Landauer70,Anderson80}.
Using scaling arguments, \cite{Anderson80} the 
ensemble-averaged resistivity of harmonic systems 
having binary isotopic 
disorder (single-valued impurities) 
can be expressed as a function of the scattering 
cross section of a single impurity and the impurity 
concentration.
Binary isotopically disodered harmonic chains also have 
the property that in the limit that the impurity 
concentration approaches unity, the system is once 
again ``pure,'' and the Anderson localization length 
diverges toward infinity.
As a result, the Anderson localization length passes through a 
minimum at intermediate concentrations \cite{Sheng86b}.
These properties of a binary isotopically disordered harmonic chain 
(BIDHC) make these systems both tractable and interesting.

In this work, an approximation is developed for the localization 
length of a BIDHC with arbitrary impurity cross section and 
concentration.
In addition, the result incorporates the discrete nature of a 
classical chain.
Similar studies have been performed previously, but with
important differences.
Bourbonnais and Maynard \cite{Bourbonnais90a} studied 
energy transport in one-dimensional 
systems having isotopic disorder, but the impurity masses
were not single-valued.
The results of Azbel and Soven \cite{Azbel82} on 
quantum systems having binary isotopic disorder
were based on a continuum solution for impurity 
locations constrained to exist on lattice sites.
Although the Azbel and Soven result 
applied to short wavelengths that may exist between 
the impurities, the results do not incorporate the additional 
features of a discrete mechanical system.

This study serves as an introduction to future work on energy 
transport in binary isotopically disordered anharmonic chains.
These anharmonic chains serve as a useful model for, among 
other things, line width broadening
\cite{Held97,Rohmfeld01,Widulle02}.
To perform that work,
numerical integration will used to study the time-dependent
nature of these systems.
To be practical, the initial conditions will require sufficiently short 
wavelengths and high impurity concentrations to keep integration 
times manageable.
In addition, the results of the study will incorporate the discrete
nature of the mechanic chains so that these effects can 
accounted for in the results.

In this study, the energy localization in a BIDHC 
is studied for arbitrary displacement wavelength, impurity concentration and 
scattering cross section.
Disorder is effected by changing randomly selected masses by a fixed amount.
A continuum Kronig-Penney (KP) model \cite{Kronig31} 
is used to develop a general expression 
for scattering cross section, and the continuum impurity impedance is 
corrected for wavelength dependent pulse propagation speed in discrete
systems.
The resulting expression is verified by direct numerical integration.
The localization length of systems with strong scatterers is calculated using 
both the continuum KP model and the MacKinnon and Kramer \cite{MacKinnon81} 
(MK) method.
The distribution of Lyapunov exponents is studied using the continuum 
KP model, and the minimum requirements are found for system length and 
number of scatterers to achieve proper scaling statistics.
The localization length concentration dependence is studied using the MK method, 
and an analytical expression is found for its behavior.
For arbitrary impurity concentration and impurity cross section,
the localization length in a BIDHC 
can be approximated by invoking a simple {\em ansatz} based on an 
analogy to electrical systems.
The result is accurate for systems having displacement wavelengths
at least four lattice spacings long.
%

%
%

\section{Model System}
The physical model used here is the harmonic version of the 
Fermi-Pasta-Ulam (FPU) \cite{Fermi55} chain that is composed of discrete springs 
and masses.
The masses $m_i$ are on a lattice with spacing $\UnitLength$ and interact via nearest 
neighbor springs with force constant $K$.
Disorder is effected by changing the background mass $\EquilMass$ by a fixed amount 
$\ImpurityMass$ with probability $c$.
To simplify the results, all lengths are scaled by the lattice spacing $\UnitLength$.

For a system composed of $N$ masses, each characterized by a displacement 
$x_i$ about the equilibrium location and a momentum $p_i$, the Hamiltonian
is separable:
\begin{equation}
   H = \frac{1}{2}\sum_i^N \frac{p_i^2}{m_i} 
	+ \SpringConstant\,\left(x_{i+1}-x_i\right)^2
   \label{eqn:Hamiltonian}
\end{equation}
The real space equation of motion is 
\begin{equation}
   \frac{m_i}{K}\,\ddot{x}_i = x_{i+1} - 2x_i + x_{i-1}
   \label{eqn:xeom}
\end{equation}
The Fourier transform leads to the corresponding equation of motion 
for the energy eigenstate amplitudes $u_i(\omega)$:
\begin{equation}
	u_{i+1} = 
		\left[2 - \frac{\omega^2 m_i}{\SpringConstant}\right]\,
		u_i - u_{i-1}
	\label{eqn:ueom}
\end{equation}
This is the corresponding Anderson tight-binding model for the 
chain.

\subsection{Discrete Analysis}

The time-dependent properties of the system were determined by numerical 
time integration of Eqn.~\ref{eqn:xeom} 
using a fourth-order symplectic integrator algorithm (SIA4) 
for separable Hamiltonians.
The coefficients were taken from Candy and Rozmus \cite{Candy91}, and
the time step was 1/200 of the natural period.


The localization length was calculated from Eqn.~\ref{eqn:ueom} using the method 
of MacKinnon and Kramer (MK) \cite{MacKinnon81}.  This method exploits the 
statistical properties of the $u_i$ so that periodic rescaling can be used to improve 
overall statistics.

%
%

\subsection{Continuum Analysis}

A Kronig-Penney model \cite{Kronig31} is used to develop an expression for 
the scattering cross section of an impurity and to study the statistics of the 
scaling parameter \cite{Anderson80}.  The continuum system analogous to the 
discrete chain is a homogeneous elastic rod having mass density 
$\MassDensity$ and Youngs modulus $\YoungsModulus$.
In the absence of impurities, a longitudinal displacement amplitude 
$\psi(x,t|\omega)$ with angular velocity $\omega$ will propagate down the rod 
with longitudinal velocity $c_l = \sqrt{\YoungsModulus/\MassDensity}$.

A harmonic oscillator impurity located at $x^\prime$ will give rise to a 
reactive force due to the impurity impedance $Z$ for a wave with
angular velocity $\omega$:
\begin{equation}
   \left[
	\MassDensity\frac{\partial^2}{\partial t^2}
	- \YoungsModulus \frac{\partial^2}{\partial x^2} = 
	-Z(\omega)\,\delta(x-x^\prime) \frac{\partial}{\partial t}
	   \right]\,\psi(x,t|\omega)
  \label{eqn:continuum}
\end{equation}
We assume the solution has a time dependence given by an exponential of 
angular velocity $\omega$ ($\psi=\phi(x|\omega)\,\E^{-\I\omega t}$) 
to obtain
\begin{equation}
   \left[\frac{\partial^2}{\partial x^2} + \kappa^2  
   	= \frac{-\I\omega}{\YoungsModulus}\,
		Z(\omega)\,\delta(x-x^\prime)
	\right] \phi(x|\omega)
   \label{eqn:Helmholtz}
\end{equation}
where $\kappa=\omega/c_l$.
Equation~\ref{eqn:Helmholtz} can be solved analytically 
for a single scatterer or can be solved numerically for a particular 
system composed of a number of 
scatterers.  
From this numerical solution, one can determine the system resistivity, which can 
be used to calculate the localization length.

%
%

\section{Cross Section}

The scattering cross section $\CrossSection$ for a single impurity can be calculated from 
the continuum system of Eqn.~\ref{eqn:Helmholtz}.
The Green's function for the 1-D Helmholtz equation \cite{Arfken70} can be used to solve for the admittance $\phi(x)$:
\begin{equation}
   \phi(x) = \E^{\I \kappa x} - 
	\left(\frac{Z(\omega)}{Z(\omega)+ \sqrt{4\MassDensity\YoungsModulus}}
		\right)\,
	\E^{\I \kappa |x-x_i|}\,
	\E^{\I \kappa x_i}
\end{equation}
The scattering cross section in 
1-D is equivalent to the reflection probability:
\begin{equation}
   \CrossSection = \frac{|Z(\omega)|^2}{|Z(\omega)|^2 
			+ 4\MassDensity\YoungsModulus}
\end{equation}
To apply this equation to the discrete chain, it must be converted from a continuum
description to the corresponding discrete description.

For an FPU system having masses spaced a distance $a$ apart, the 
continuum coefficients can be expressed as a function of the discrete 
properties in the limit $a\rightarrow 0$ \cite{Goldstein58}:
\begin{equation}
   \MassDensity = \frac{\EquilMass}{a}
	\hspace{1.0in}
   \YoungsModulus = \SpringConstant a
\end{equation}
The scattering cross section can now be expressed as a function of the 
discrete system components:
\begin{equation}
   \CrossSection = \frac{|Z(\omega)|^2}{|Z(\omega)|^2 
			+ 4\SpringConstant\EquilMass}
   \label{sigma:eqn}
\end{equation}

The impedance $Z(\omega)$ 
of a mass impurity along a one-dimensional chain 
is proportional to the mass $\ImpurityMass$ 
that is added to the background mass $\EquilMass$
\cite{Morse68}:
\begin{equation}
   Z(\omega) = -\I\,\omega \ImpurityMass / \PulseSpeed
   \label{Z:eqn}
\end{equation}
This equation has been modified from the continuum relation to account 
for the properties of a discrete one-dimensional chain.  A displacement 
wavelength has a corresponding wave number
$\WaveNumber = 2\pi/\Wavelength$.  For the discrete chain, the 
angular velocity $\omega$ and relative pulse propagation 
speed $\PulseSpeed$ can be expressed as a function of the 
displacement wave number $\WaveNumber$ \cite{Kittel86}:
\begin{equation}
   \omega = 2\sin(\WaveNumber/2)
   	\hspace{2cm}
   \PulseSpeed = \cos(\WaveNumber/2)
   \label{c_s:eqn}
\end{equation}
Finally, substituting Eqn.~\ref{Z:eqn} into 
Eqn.~\ref{sigma:eqn} gives the scattering cross section of a 
mass impurity in a one-dimensional chain:
\begin{equation}
   \CrossSection = \frac{(\ImpurityMass\, \omega / \PulseSpeed)^2}%
		{(\ImpurityMass\, \omega / \PulseSpeed)^2 + 4\SpringConstant\EquilMass}
   \label{cs:eqn}
\end{equation}

\begin{figure}
  \includegraphics[angle=0,width=\columnwidth]{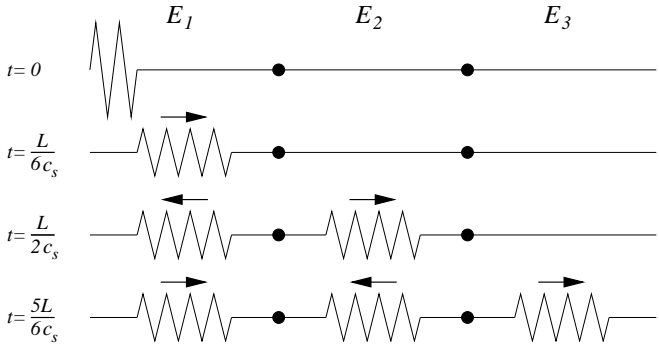}
  \caption{Schematic of cross section numerical experiment;
	each line represents the state of the system at the time of
	measurement.
	The time $t$, shown along left side, is expressed as a function 
	of the system length $L$ and pulse propagation speed $\PulseSpeed$.
	The large dots denote the location of two impurities.}
  \label{cs_schem:fig}
\end{figure}

A numerical experiment was performed to test the applicability of 
Eqn.~\ref{cs:eqn} to FPU systems, and a schematic of the experiment 
is shown in Fig.~\ref{cs_schem:fig}.  The system had fixed ends and length 
$L$.  One impurity was located at $L/3$, and another impurity was 
located at $2L/3$.  
An initial displacement  was made with wavelength $\Wavelength$ and total 
length 16$\Wavelength$, and the hyperbolic tangent function 
was used to taper the oscillation amplitude from $A$ to zero.  Initially, the 
pulse was located at one end of the system.  It had length 8$\Wavelength$, 
amplitude 2$A$, and zero initial velocity.  The time-dependent behavior was 
determined by numerical integration using the SIA4 algorithm.

The measured scattering 
cross section was determined from the energy located in the 
three regions separated by the impurities.    
The pulse energy located in each three regions, denoted by 
$E_1$, $E_2$, and $E_3$,  was calculated at 
four separate times: $t=0$, $L/6 \PulseSpeed$, $L/2 \PulseSpeed$, and 
$5L/6 \PulseSpeed$.
At these times, the pulses are located near the center of a 
region.  Although redundant with respect to the initial total energy, 
determining the energy at $t=L/6 \PulseSpeed$ 
provided a consistency check.  In each case, the difference between 
$E_1(t=0)$ and $E_1(t=L/6\PulseSpeed)$ was less than one part in 10$^4$.

The scattering cross section in FPU systems is the reflection coefficient.
In this experiment,
the reflection coefficient $R$ is calculated from the ratio of energies in the first 
two intervals after one scattering event: 
\begin{equation}
   R = \frac{E_1(L/2\PulseSpeed)}{E_1(0)}
   \label{R:eqn}
\end{equation}
Although the transmission coefficient $T$ could have been determined from 
$E_2(L/2\PulseSpeed)$, a second scatterer was used as a more rigorous test 
of the experiment design and numerical integrator.
The transmission coefficient $T$ for a single scatterer was 
calculated using the energy after two scattering events:
\begin{equation}
   T = \sqrt{\frac{E_3(5L/6\PulseSpeed)}{E_1(0)}}
\end{equation}
In all cases, the magnitude of $1-R-T$ was less than 10$^{-3}$.

\begin{figure}
  \includegraphics[width=\columnwidth]{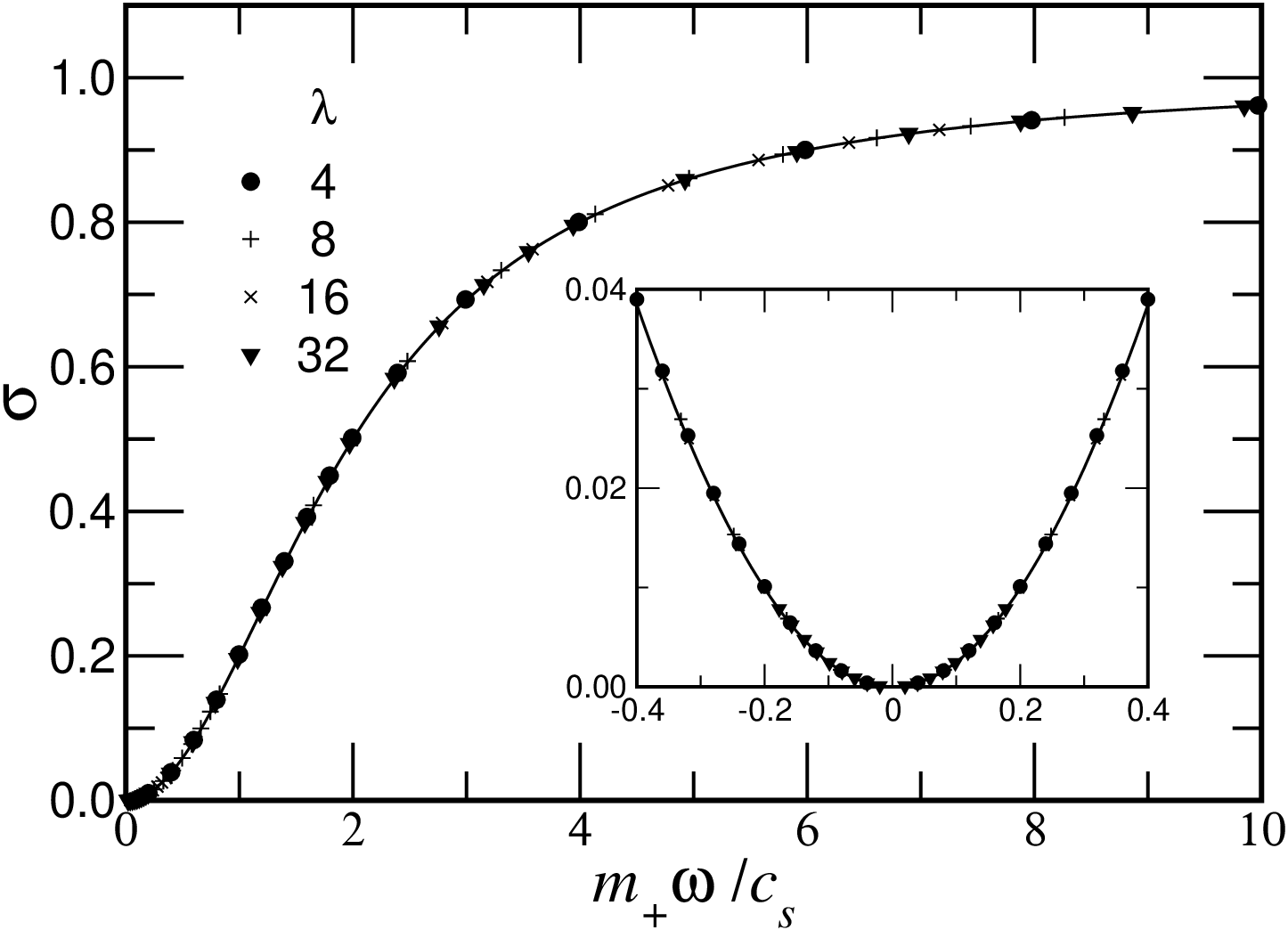}
  \caption{Cross section $\CrossSection$ as a function of  
	$\ImpurityMass\omega/\PulseSpeed$ for different wavelengths $\Wavelength$.  	
	The solid curve is the analytical result in 
	Eqn.~\protect{\ref{cs:eqn}}.
	The inset shows additional data near zero.}
  \label{cs:fig}
\end{figure}

A comparison of the estimated cross section $\CrossSection$ 
in Eqn.~\ref{cs:eqn} 
to the measured reflection coefficient $R$ in Eqn.~\ref{R:eqn} is 
shown in Fig.~\ref{cs:fig} for different displacement wavelengths 
$\Wavelength$.  The results demonstrate that Eqn.~\ref{cs:eqn} is
an accurate estimate for the scattering cross section $\CrossSection$
for displacement wavelengths as short as 4. 
Moreover, the symmetry about zero for scattering cross section for negative 
values of $\ImpurityMass$ is shown in the inset of Fig.~\ref{cs:fig}.

%
%

\section{Localization Length}

The resistivity $\Resistivity_N$ of a system
composed of $N$ scatterers is 
the ratio of the system reflection coefficient $R_N$ to the system 
transmission coefficient $T_N$ 
\cite{Landauer70,Anderson80}:
\begin{equation}
	\Resistivity_N = \frac{R_N}{T_N}
	\label{eqn:Resistivity}
\end{equation}
The localization length of the system is defined in an averaged sense.  To achieve proper scaling, 
however, one defines its inverse, the Lyapunov exponent 
$\LExp$:
\begin{equation}
   \LExp = \frac{\ln\left(1+\Resistivity_N\right)}{L}
\end{equation}
More specifically, there is a distribution of values from an ensemble of systems.  If the systems 
composing the ensemble are sufficiently large, the distribution of $\LExp$ values will be normal.
The localization length $\LocLength$ for a system having length $L$ and $N$ scatterers is 
defined from the ensemble averaged Lyapunov exponent:
\begin{equation}
   \LocLength^{-1} = \langle\LExp\rangle 
         = \frac{\langle\ln\left(1+\Resistivity_N\right)\rangle}{L}
   \label{LocLength:eqn}
\end{equation}
Unless otherwise noted, the symbol $\LExp$ shall imply the ensemble averaged quantity.

%
%

\subsection{Strong Scatterers}

To perform numerical experiments on anharmonic systems of manageable length, the scatters 
will need to be relatively strong.  Therefore, weak scattering results will not be applicable.
Moreover, a means is needed to predict the averaged localization behavior of a system 
using only single scatterer information. 
Because the impurities are identical, the scaling law \cite{Anderson80} can be 
exploited to express system behavior as a function of 
the resistivity of a single scatterer $\Resistivity$:
\begin{equation}
	\langle\ln(1+\Resistivity_N)\rangle  = N\ln(1+\Resistivity)
	\label{Nln:eqn}
\end{equation}
Substituting from Eqn.~\ref{LocLength:eqn} above yields an unbiased estimate for the 
ensemble averaged localization length (dilute limite) as a function of impurity concentration $c$:
\begin{equation}
   \LocLengthDL^{-1} = c \langle\ln(1+\Resistivity)\rangle
   \label{eqn:LocLength}
\end{equation}
In the limit of weak scattering ($c,\Resistivity\rightarrow 0$), 
one recovers the expected result 
$\LocLengthDL^{-1} = c\CrossSection = \MFP^{-1}$, where $\MFP$ is the 
classical mean free path.

\begin{figure}
  \includegraphics[width=\columnwidth]{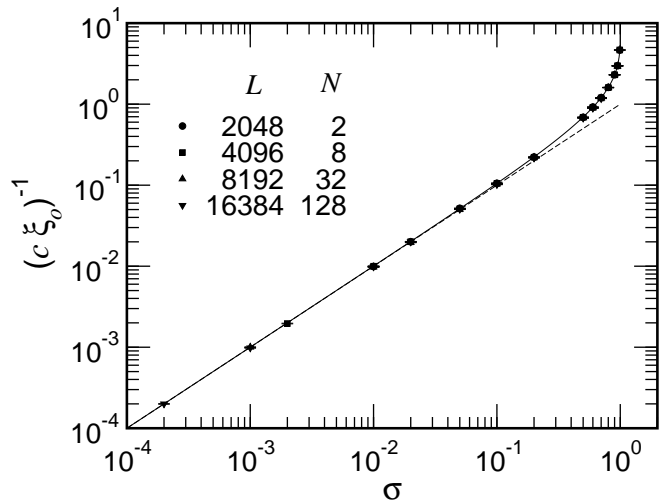}
  \caption{Localization length $\LocLengthDL$, scaled by impurity concentration $c$, as a 
       function of impurity scattering cross section $\CrossSection$ for systems having length
       $L$ and $N$ impurities.
       The solid line is $\ln(1+\Resistivity)$ and the dashed line represents the mean free 
       path estimate.
       The error bars represent the mean standard deviation.
       (Many of the symbols lie upon one another and the error bars are 
	typically smaller than the symbols.)}
  \label{xiN:fig}
\end{figure}

To demonstrate both the effect of strong scatterers ($\CrossSection\rightarrow 1$) 
and the accuracy of Eqn.~\ref{Nln:eqn},
the ensemble average $\langle\ln(1+\Resistivity_N)\rangle$ was calculated 
from 10~000 systems, each having length $L$ and $N$ impurities.  The 
displacement wavelength 
was 32, and the scattering cross section of each impurity was $\CrossSection$.
The results from the calculation are shown in Fig.~\ref{xiN:fig} as a function of the 
impurity cross section $\CrossSection$.
In the figure, the error bars represent the mean standard deviation, and many of the 
symbols lie upon one another.
As can be seen, all the systems have the $\ln(1+\Resistivity)$ dependence that 
deviates from the mean free path approximation ($\LocLengthDL^{-1}=c\CrossSection$) 
for cross sections greater than approximately 0.2 for all combinations of system size and 
number of scatterers.

%
%

\subsection{Statistics}

The results shown in Fig.~\ref{xiN:fig} demonstrate that, in the mean, 
systems having a finite density 
of scatterers have the expected behavior.  Recent results 
suggest that systems 
having relatively few scatterers do not exhibit Gaussian behavior and, therefore, not 
obey scaling laws.
For systems having sufficient length and number of scatterers,
the population of Lyapunov exponents is normally 
distributed \cite{Anderson80}, with variance
$\VAR{\LExp}$\,\, \cite{Anderson80,Deych01}
\begin{equation}
	\VAR{\LExp} = \frac{2}{L^2}\ 
		\langle\ln(1+\Resistivity_N)\rangle
	\label{variance:eqn}
\end{equation}

Returning to the data of Fig.~\ref{xiN:fig}, the population of Lyapunov exponents $\LExp$ 
was compared to the expectations of Eqn.~\ref{variance:eqn}.  To assess the ``normality'' 
of the data, the intervals, both above and below the mean, having coverage factors 
\cite{BIPM} 
corresponding to one and two standard deviations were determined from the population.
In addition, the population standard deviation $\Moment$ so that it could be compared
to both its estimated value in Eqn.~\ref{variance:eqn} and to the corresponding coverage interval.

\begin{figure}
  \includegraphics[width=\columnwidth]{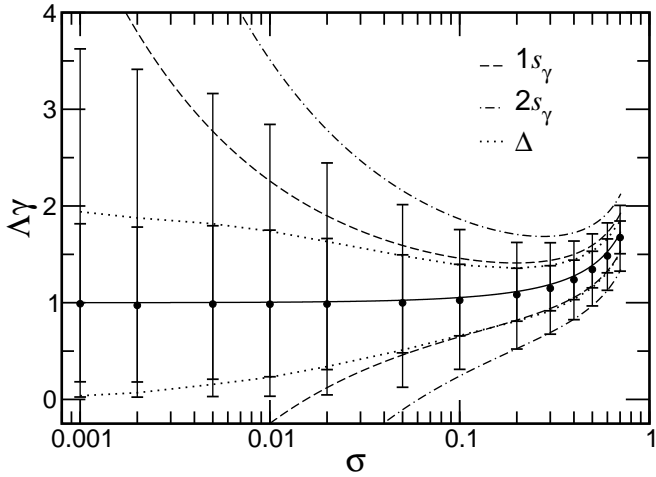}
  \caption{Lyapunov exponent $\LExp$ statistics as a function of 
	scattering cross section $\CrossSection$ for a system 
	with $L=16~384$, $N=128$, and $\Wavelength=32$.
	Error bars represent coverage factors corresponding to one and two 
	standard deviations of a normal distribution. 	
	The predicted intervals $\pm 1\SD{\LExp}$ and $\pm 2\SD{\LExp}$ 
	(Eqn.~\protect{\ref{variance:eqn}}) 
	are shown, along with observed population standard deviation $\Moment$.}
  \label{stat:fig}
\end{figure}

The results of the Lyapunov exponent statistics calculation, 
from an ensemble with a population of 10~000, are shown in Fig.~\ref{stat:fig} for 
systems having length $L=16~384$, $N=128$ impurities, and displacement 
wavelength $\Wavelength=32$.  (Results for other systems were similar, and are 
omitted for brevity.)  The data are shown as a function of the single impurity 
cross section $\CrossSection$.
The filled symbols are the average value, and the solid line is the estimated average value.
The error bars represent the intervals that have the same coverage factor as one and 
two standard deviations in a normal distribution.  The two pairs of dashed lines are the estimated standard deviations from Eqn.~\ref{variance:eqn}.
The dotted line is the population standard deviation $\Moment$.

For small scattering cross sections, the distribution of $\LExp$ is asymmetric, with zero as a 
lower bound for the coverage intervals.  Interestingly, only the outer intervals are asymmetric 
about the mean.  The inner intervals are nearly symmetric about the mean, and they 
have a value nearly equal to the population standard deviation.

As the scattering cross section increases, the results begin to exhibit Gaussian 
behavior.  Above a 0.2 scattering cross section,  the measured intervals, the predicted intervals 
$\SD{\LExp}$, and the population standard deviation $\Moment$ all agree.  Therefore, 
agreement between the population standard deviation $\Moment$ and the estimated standard deviation $\SD{\LExp}$ is as much as measure 
of ``normality'' as is a careful analysis of the population coverage intervals.

\begin{figure}
   \includegraphics[width=\columnwidth]{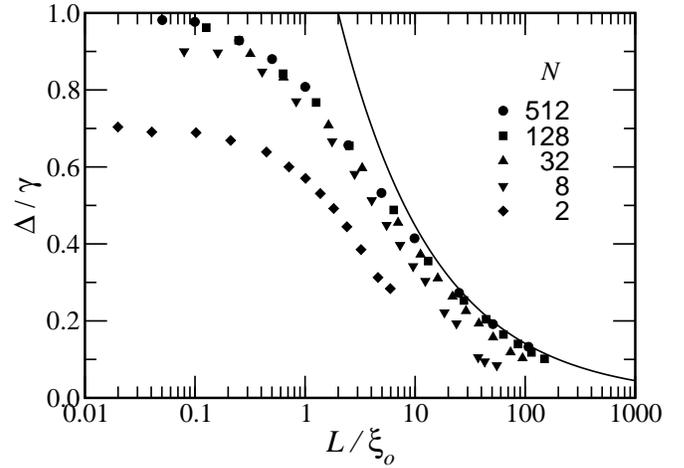}
   \caption{Lyapunov exponent coefficient of variation $\Moment/\LExp$ as a 
	function of the ratio of system length $L$ to the localization
	length $\LocLengthDL$ for systems having different number of 
	scatterers $N$.  The solid curve is Eqn.~\protect{\ref{cov:eqn}}.}
   \label{cov:fig}
\end{figure}

Exploiting this relationship, the population standard deviation $\Moment$ for 
all the data from the Lyapunov exponent statistics experiment 
are shown in Fig.~\ref{cov:fig} as a function of the system length.  For normally distributed 
populations, the estimated coefficient of variation can be determined from 
Eqn.~\ref{LocLength:eqn} and Eqn.~\ref{variance:eqn}, and has the following convenient form:
\begin{equation}
   \frac{\SD{\LExp}}{\langle\LExp\rangle} 
   	= \sqrt{\frac{2}{L/\LocLengthDL}}
  \label{cov:eqn}
\end{equation}
This equation is shown as a solid line in Fig.~\ref{cov:fig}.
One can conclude from the figure that in order for the observed Lyapunov exponents to be 
normally distributed, the system length must be at least 10 times the localization length, and 
the number of scatterers must be greater than approximately 32.

%
%

\section{Concentrated Impurities}

For dilute impurity concentrations, the localization length decreases with 
increasing impurity concentration.
As the impurity concentration $c$ approaches unity, however, 
the FPU system will become a pure system composed entirely of masses 
$\EquilMass + \ImpurityMass$.  At $c=1$, the system is once again devoid of 
impurity and the localization length goes to infinity.
At intermediate impurity concentrations, the localization length passes
through a minimum.  Therefore, the behavior of a system at arbitrary impurity 
concentration cannot be fully characterized by the relation in Eqn.~\ref{eqn:LocLength}.

For dilute systems, the localization length $\LocLength_{c\rightarrow 0}$ is as before:
\begin{equation}
	\LocLength_{c\rightarrow 0}^{-1} 
	= \LocLengthDL^{-1} = c \, \langle\ln (1+\Resistivity)\rangle
\end{equation}
At high concentrations, the localization length $\LocLength_{c\rightarrow 1}$ has a 
concentration dependence that is proportional to $(1-c)$:
\begin{equation}
	\LocLength_{c\rightarrow 1}^{-1} = (1-c)\langle\ln (1+\Resistivity^\prime)\rangle
\end{equation}
The adjusted resistivity $\Resistivity^\prime$ is for a system having 
background mass $\EquilMass+\ImpurityMass$ and impurities with 
mass $\EquilMass$.  
By the nature of the solution using the MK method, and given that the systems 
are harmonic, the frequency $\omega$ is 
the same for both systems, but the wavelength and the corresponding 
pulse speed are different:
\begin{eqnarray}
	\WaveNumber^\prime	& = & \WaveNumber\,\sqrt{\EquilMass+\ImpurityMass}\\
	\PulseSpeed^\prime 	& = & \cos\left(\WaveNumber^\prime/2\right) \\
	\Resistivity^\prime & = & \frac{(-\ImpurityMass\omega/\PulseSpeed^\prime)^2}{%
		4\SpringConstant(\EquilMass+\ImpurityMass)} 
\end{eqnarray}

The behavior of the system for all values of impurity concentration 
is conjectured from the electrical analogy: as Lyapunov exponent is to 
resistivity, localization length is to conductivity.  If one assumes that 
at some intermediate concentration the behavior is simultaneously 
expressing itself as two systems with localization lengths 
$\LocLength_{c\rightarrow 0}$ and $\LocLength_{c\rightarrow 1}$, 
these two systems should contribute 
independently to the overall behavior.  By analogy to conductors,
the total localization length $\LocLength$ would be additive:
\begin{equation}
	\LocLength = \LocLength_{c\rightarrow 0} \,+\, \LocLength_{c\rightarrow 1}
	\label{loc_total:eqn}
\end{equation}
This equation represents a more complete estimate for the localization length that is 
valid for both strong and concentrated scatterers.

%
%

\subsection{Positive $\ImpurityMass$}

\begin{figure}
  \includegraphics[width=\columnwidth]{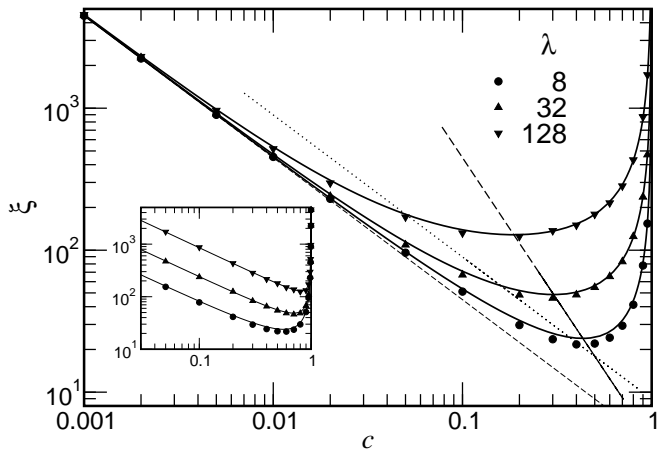}
  \caption{Localization length $\LocLength$ as a function of impurity 
	concentration $c$ in a discrete system having scatterers with cross section 
	0.2.
	Solid lines is Eqn.~\protect{\ref{loc_total:eqn}}, 
	dashed line is locus of minima, and
	dotted line denotes equal contribution from $\LocLength_{c\rightarrow 0}$ and 
	$\LocLength_{c\rightarrow 1}$.
	Inset shows same data as a function of (1-$c$).}
  \label{phase:fig}
\end{figure}

The accuracy of this approximation is shown in Fig.~\ref{phase:fig} for 
systems 
having impurity scattering cross section $\CrossSection=0.20$ and displacement
wavelengths ranging from 8 to 128.  The localization length $\LocLength$ 
for these systems was calculated using the MK solution.  As can be seen, 
the approximation in Eqn.~\ref{loc_total:eqn} is reasonably accurate.
The inset shows the same data, plotted as a function of $(1-c)$, highlighting 
the separate behavior near $c\rightarrow 1$ for the different wavelengths.

Also shown in Fig.~\ref{phase:fig} are a dashed line and a dotted line.  The 
dashed line is the locus of localization length minima as a function of 
displacement wavelength.  Not only does the minimum localization length 
increase with increasing wavelength, the concentration at which this happens 
decreases with increasing wavelength.  The dotted line in Fig.~\ref{phase:fig}
is the locus of points where 
$\LocLength_{c\rightarrow 0}=\LocLength_{c\rightarrow 1}$.  
This locus of points is meant to delineate the 
cross-over point as the system passes from one dominate phase to the other.
The cross-over point has a stronger concentration dependence than the locus 
of minima.

%
%

\subsection{Negative $\ImpurityMass$}

\begin{figure}
  \includegraphics[width=\columnwidth]{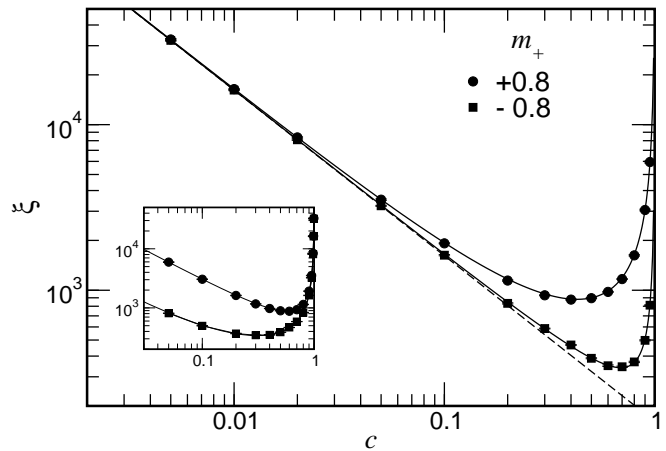}
  \caption{Localization length $\LocLength$ as a function of impurity 
	concentration $c$ for $\ImpurityMass = \pm$ 0.8. 
	Inset shows same data as a function of (1-$c$).}
  \label{neg_cs_phase:fig}
\end{figure}

At a given oscillation wavelength and frequency, 
the impurity scattering cross section depends upon the magnitude of the 
change in mass $\ImpurityMass$.  For small impurity concentrations, the 
localization length for a system with impurity mass $\EquilMass+\ImpurityMass$ will be
equal to one with impurity mass $\EquilMass-\ImpurityMass$.  At higher concentrations, 
however, the behavior of systems will differ.  

As an example, the localization concentration dependence was calculated for two systems 
with displacement wavelength $\Wavelength=32$.  The added impurity masses were 
$+0.8$ and $-0.8$, and the localization length was determined by the MK method.
The results of the calculation are shown in Fig.~\ref{neg_cs_phase:fig} as a function of the impurity concentration $c$.  
The estimate from Eqn.~\ref{loc_total:eqn} is shown as a solid line.
Within the inset are the data plotted as a function of $(1-c)$.
As expected, the behavior of the two systems diverge at higher concentrations.


%
%
\subsection{Azbel and Soven Comparison}

\begin{figure}
  \includegraphics[width=\columnwidth]{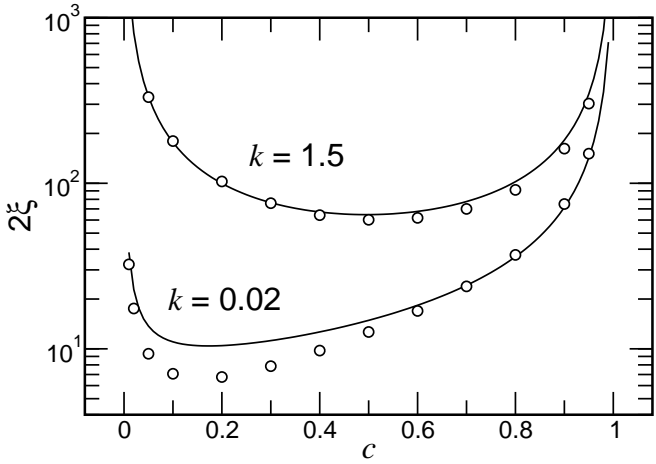}
  \caption{Localization length $\LocLength$ as a function of 
  	impurity concentration $c$ for two systems with 
	different wavenumber $\WaveNumber$.
	Open symbols are calculations using the MK method, 
	and solid lines are Eqn.~\protect\ref{loc_total:eqn}.
	The figure can be compared directly to 
	Figure 1 of Azbel and Soven\cite{Azbel82}}
  \label{AS:fig}
\end{figure}

To better judge performance, Eqn.~\ref{loc_total:eqn} is 
compared to the more rigorous result of Azbel and Soven 
\cite{Azbel82} (AS).
The AS model contains quantum particles interacting 
with delta function potentials that have strength $V$ 
and are located at random integer values of $x$.
For the AS systems, the value of $V$ is equal to -1, and 
the scattering cross section 
$\CrossSection_{\textrm{AS}}$ of each scatterer is 
a function of the particle wave number $\WaveNumber$:
\begin{equation}
	\CrossSection_{\textrm{AS}} = \frac{1}{1+16 \sin^2 (\WaveNumber/2)}
\end{equation}
This is sufficient to duplicate the AS numerical calculation.

Figure 1 of Azbel and Soven\cite{Azbel82} shows results from calculations 
made for three values of wavenumber $\WaveNumber$: 0.02, 1.5, 3.13.
The MK method is used here to duplicate the numerical results for the two smaller values 
of $\WaveNumber$, and the results are shown as open symbols in 
Fig.~\ref{AS:fig}.  
The solid lines in Fig.~\ref{AS:fig} are the corresponding estimate of 
Eqn.~\ref{loc_total:eqn}.
Two things should be noted explicitly:
The AS definition of localization length corresponds to 
twice the localization length defined here.
The definition of localization length in Azbel and Soven\cite{Azbel82} uses 
the Landauer \cite{Landauer70} scaling parameter, while subsequent
work \cite{Azbel83a,Azbel83b} use the scaling of Anderson et al.~\cite{Anderson80}.
The results shown in Fig.~\ref{AS:fig} use the latter scaling definitions.

%
%

\subsection{$c\Wavelength$ Effect}

\begin{figure}
  \includegraphics[width=\columnwidth]{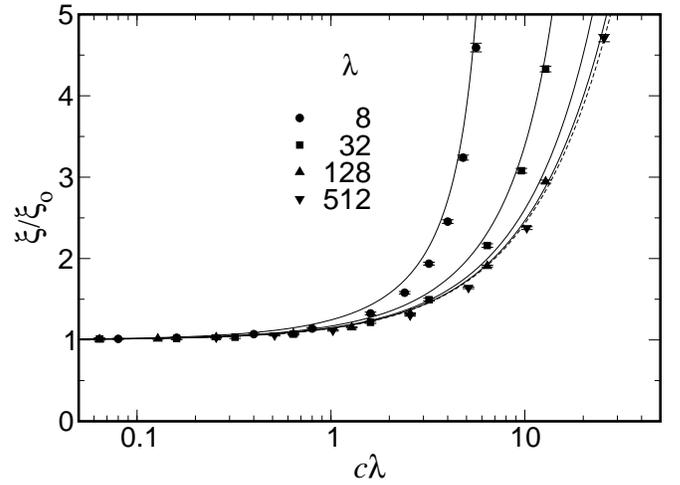}
  \caption{The ratio of the measured localization $\LocLength$ to the 
	calculated localization $\LocLength_\circ$ as a function of 
	product of impurity concentration $c$ and wavelength $\Wavelength$.
	Filled symbols are calculated solutions for systems having impurity 
	cross section 0.2. 
	The dotted line is the limiting curve for $\Wavelength\rightarrow\infty$.}
  \label{clambda:fig}
\end{figure}

As can be seen in Figs.~\ref{phase:fig} and 
\ref{neg_cs_phase:fig}, the behavior 
of the total localization length diverges from the dilute limit $\LocLengthDL$.
In Fig.~\ref{phase:fig}, the point at which $\LocLength$ begins to differ from 
$\LocLengthDL$ is a function of the displacement wavelength.  To more 
clearly demonstrate this effect, 
the ratio $\LocLength/\LocLengthDL$ is shown in Fig.~\ref{clambda:fig} 
as a function of the product $c\Wavelength$.  The data shown are those 
appearing in Fig.~\ref{phase:fig}, with the addition of those 
for $\Wavelength=512$.  Also shown in the figure is a dotted line denoting 
the long wavelength limiting behavior 
for $\Wavelength\rightarrow\infty$.
For values of $c\Wavelength$ greater than 1, the dilute limit approximation 
does not hold, and the observed localization length is greater than the 
dilute limit estimation.  
Therefore,
even though the effect is less dramatic for increasing $\Wavelength$, there 
is a minimum effect, regardless of the wavelength.

%
%

\section{Anharmonic Chains}

\begin{figure}
  \includegraphics[width=\columnwidth]{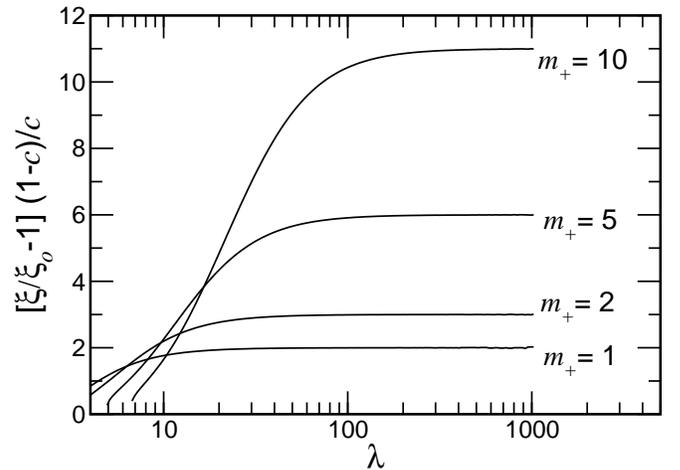}
  \caption{The ratio of the measured localization $\LocLength$ to the 
	dilute limit localization length $\LocLength_\circ$, normalized by the 
	impurity concentration $c$, as a function of 
	displacement wavelength $\Wavelength$.}
  \label{xi_lambda:fig}
\end{figure}

In an anharmonic system, phonon-phonon interactions will 
lead to the creation of displacements with varying wavelength.  
In time, very long wavelength displacements will be created.
Because the 
impurity masses are fixed in time, the scattering cross section 
for the displacement waves will decrease as $\Wavelength^{-2}$.  In addition, 
because the impurities are fixed in space, the product $c\Wavelength$ 
will increase over time.  Based on the results from the last Section, one would expect that 
the ratio $\LocLength/\LocLengthDL$ will diverge to infinity with increasing wavelength.

Using Eqn.~\ref{loc_total:eqn}, the ratio $\LocLength/\LocLengthDL$ was calculated 
as a function of displacement wavelength for systems having constant impurity mass and 
concentration.  The results are shown in Fig.~\ref{xi_lambda:fig} for different values for 
$\ImpurityMass$.  For all values of impurity concentration $c$, the ratio 
$\LocLength/\LocLengthDL$ asymptotes to a constant at long wavelength:
\begin{equation}
   \lim_{\Wavelength\rightarrow\infty} \frac{\LocLength}{\LocLengthDL} =
   	1 + \frac{c}{1-c}\,\left(\EquilMass+\ImpurityMass\right)
   \label{xi_lambda:eqn}
\end{equation}
Therefore, for anharmonic systems, the localization length is characterized, to within 
a constant, by the dilute limit expression for $\LocLengthDL$ given in 
Eqn.~\ref{eqn:LocLength}.

The same relationship applies to both positive and negative values for 
$\ImpurityMass$.  For negative values of $\ImpurityMass$, in fact, in the limit 
$\ImpurityMass\rightarrow -1$, the 
behavior of the system is accurately characterized 
by $\LocLengthDL$ at long wavelengths.

\section{Conclusion}

The localization length for harmonic chains having binary disorder can be predicted accurately 
over a wide range of wavelengths, 
impurity cross section, and impurity concentration.
The ingredient needed for this prediction is the cross section 
of a single scatterer, corrected for short wavelength displacements via the relative 
pulse propagation speed.
The localization length over the entire impurity concentration range can then be estimated by 
approximating the system as a sum of two independent systems, each accounting for the 
behavior of the system at the two limits of impurity concentration.

The general result applies to systems in which the impurity mass is either larger or smaller 
that the original mass.
Although the scattering cross section is symmetric about zero, with respect to the mass 
added to the background value, the localization length behavior differs for negative and 
positive changes in mass having the same scattering cross section.
This difference in behavior with respect to localization length, along with previous results 
showing differences in the rate of phonon-phonon interactions, are discussed in the 
context of numerical experiments on anharmonic systems.

The general results also suggest that the localization length of long 
displacement wavelengths created by phonon-phonon interactions can be 
approximated, to within a constant, from dilute limit calculations results.
For increasing concentration 
and constant impurity concentration, the localization length, with respect to the dilute limit 
prediction, will eventually diverge toward infinity.
This deviation occurs for all wavelengths, and is a universal function of the impurity 
concentration and the displacement wavelength, for a constant cross section.
By contrast, 
in a numerical experiment on an anharmonic system composed of fixed scatterers, the 
scattering cross section decreases with increasing wavelength.
For these systems, the long wavelength behavior is, to within a constant, accurately predicted 
by dilute limit predictions.
This conjecture will be confirmed in a future paper \cite{Snyder04}.

\begin{acknowledgments}
This work was supported by the NSF under grant number DMR-01-32726.
\end{acknowledgments}

\bibliography{/Users/jackal/jdb/j_abbrev,references}
 
\end{document}